\documentclass[a4paper,fleqn,usenatbib]{mnras}
\usepackage[T1]{fontenc}
\usepackage{ae,aecompl}
\usepackage{amstext}
\usepackage{url}
\usepackage{graphicx}	% Including figure files
\usepackage{amsmath}	% Advanced maths commands
\usepackage{amssymb}	% Extra maths symbols

\newcommand{\bea}{\begin{eqnarray}}
\newcommand{\eea}{\end{eqnarray}}
\newcommand{\be}{\begin{equation}}
\newcommand{\ee}{\end{equation}}
\newcommand{\rund}[1]{\left(#1\right)}

\newcommand{\eck}[1]{\left[ #1 \right]}

\def\elabel#1{\label{eq:#1}}

\title[Plasma lensing]{Two families of astrophysical diverging lens models}
\author[Er and Rogers]
	{Xinzhong Er$^1$ \thanks{E-mail: phioen@163.com},
	Adam Rogers$^2$ \thanks{E-mail: rogers@physics.umanitoba.ca}\\
$^1$ I.N.A.F. - Osservatorio Astronomico di Roma, via Frascati 33, 00040 - Monte Porzio Catone, Roma, Italy\\
$^2$ Department of Physics and Astronomy, University of Manitoba, Winnipeg R3T 2N2, Canada\\
}

\date{Accepted XXX. Received YYY; in original form \today}
\pubyear{2017}

\begin{document}
\label{firstpage}
\maketitle

\begin{abstract}
In the standard gravitational lensing scenario, rays from a background source are bent in the direction of a foreground lensing mass distribution.
Diverging lens behaviour produces deflections in the opposite sense to gravitational lensing, and is also of astrophysical interest.
In fact, diverging lensing due to compact distributions of plasma has been proposed as an explanation for the extreme scattering events (ESEs) that produce frequency-dependent dimming of extra-galactic radio sources, and may also be related to the refractive radio-wave phenomena observed to affect the flux density of pulsars.
In this work we study the behaviour of two families of astrophysical diverging lenses in the geometric
optics limit, the power-law and the exponential plasma lenses. Generally, the members of these model
families show distinct behaviour in terms of image formation and magnification, however the inclusion
of a finite core for certain power-law lenses can produce a caustic and critical curve morphology that
is similar to the well-studied Gaussian plasma lens.
Both model families can produce dual radial critical curves, a novel distinction from the tangential distortion usually produced by gravitational (converging) lenses. The deflection angle and magnification of a plasma lens varies with the observational frequency, producing wavelength-dependent magnifications that alter the amplitudes and the shape of the light curves. Thus, multi-wavelength observations can be used to physically constrain the distribution of the electron density in such lenses.
\end{abstract}


\section{Introduction}
\label{sec:intro}

Gravitational lensing by a mass distribution bends the path of electromagnetic radiation toward the center of the lens. The gravitational field acts like a convex lens, providing converging lens behaviour and leading to the formation of one or more images \citep{peters, SEF}. The lensing process preserves the surface brightness of a source due to Liouville's theorem, but changes the apparent solid angle of the source. This results in a net magnification of background sources \citep{narayan}. In addition to lensing by mass distributions, it is also possible for radiation to undergo diverging lensing in astronomical cases. For example, frequency-dependent lensing behaviour due to plasma has been used to study the trajectories of electromagnetic radiation in the environments around massive compact objects under the influence of both gravitation and a surrounding medium \citep{perlickBook, review, erMao, rogers15, rogers17a, rogers17b}.

Neglecting any lensing due to gravitation, plasma in the interstellar medium (ISM) can independently act as an optical medium, refracting electromagnetic rays in analogy with a conventional concave lens. Diverging lensing effects due to inhomogeneities in the ISM have been postulated as explanations for observations of demagnification of low-frequency radiation from background sources. These ``extreme scattering events'' \citep[ESEs;][]{ESE0, ESEIntro1} are also known to display dispersion \citep{romani87, cordesRickett1998, coles2010, coles2015, shannonCordes2017}. The first detected ESE occurred in observations of the quasar 0954+654 \citep{ESE0}, which underwent a frequency-dependent scattering event in 1981, showing a variation in flux density at $2.7$ GHz and $8.1$ GHz. The $2.7$ GHz observations of this event show a flux density minimum that is relatively featureless, while the corresponding minimum at $8.1$ GHz features irregular fluctuations. These fluctuations may be due to either inhomogeneities of the occulting lens or small-scale unresolved structure in the background source \citep{ESE0}.

Modeling the 0954+654 ESE using a discrete plasma structure with a Gaussian column density profile required a peak electron density of $10^5$ cm$^{-3}$ \citep{cleggFey}. Since this initial event, a handful of ESEs have been observed \citep[for example,][]{F94, ESE1}, including a detection in real time from the radio source PKS 1939-315 using the Australia Telescope Compact Array and monitored at a range of frequencies between $2$ and $11$ GHz \citep[e.g.][]{ESE3}. This ESE required a lens with electron density on the order of $10^3$ cm$^{-3}$. Assuming a thermally ionized plasma with temperature $T=10^3$ K, both of these events require local pressures in excess of the average pressure in the ISM \citep[$\sim 3\times10^3$ K cm$^{-3}$][]{ISMpressure} by a factor of $10^3$ to $10^6$. In addition to the propagation phenomena affecting the light curves of extragalactic radio sources \citep{tuntsov17}, pulsars have also been observed to display frequency dependent variations in flux density. While it is not yet clear that the nature of these events and the ESEs affecting extra-galactic radio sources are related, it is known that radio scintillation is strongly affected by refraction from local inhomogeneities in the ISM \citep{stinebring, cordesRickett1998}. The difficulties with such high pressures on the relatively small ($\sim$ AU) scales of diverging lenses can be alleviated if sheet-like plasma structures aligned along the line of sight are considered \citep{plasmaSheets}. \citet{lens1} have suggested that sheet-like electron underdensities aligned along the line of sight may solve the problem of extreme local pressures in plasma lenses. In this scenario, a source moving parallel to the short axis of an underdense plasma sheet experiences converging lensing, crosses two caustics and produces a magnification profile with a central minimum flanked by magnification peaks. In this way, \citep{lens1} can effectively reproduce the effect of a diverging lens while avoiding the problem of pressure confining the lens.

While diverging lensing is usually invoked to explain demagnification, frequency-independent events such as the demagnification of the BL-Lac type blazar J1415+1320 showed a U-shaped dimming uncorrelated with frequency between $15$ and $234$ GHz \citep{newESE2017}. The nature of these ``symmetric achromatic variabilities'' (SAVs) is unclear, and may not require plasma lensing at all. \citet{SAV} have invoked an explanation using gravitational lensing from a caustic network of extragalactic black holes. Therefore, understanding the degeneracy in the behaviour of lensing models, both diverging and converging, is an important step in uncovering the nature of such events.

Our general focus in this work is on electron overdense models which act as diverging lenses. Following previous work \citep{cleggFey, lens1, ESE2, ESE4, ESE5}, we use the gravitational lens approach to study image formation and the magnification properties of specific lens models in the geometric optics limit \citep{SEF}. We outline the basics of the formalism in Section \ref{sec:lensTheory} and in Section \ref{sec:divergingLens} we describe the simplest case of a diverging lens in the absence of any dispersive effects. We develop two families of models for electron density, the power-law profiles in Section \ref{sec:power-law}, and the exponential lenses, including the well-studied Gaussian lens, in Section \ref{sec:exponentialLens}.
We discuss our results in Section \ref{sec:discussion} and summarize our conclusions in Section \ref{sec:conclusion}.

\section{Basic formulae}
\label{sec:lensTheory}

The description of gravitational lensing used in this work follows from \citet{SEF} and \citet{narayan}. In the geometric optics limit, we adopt the thin lens approximation, which allows us to define the source plane and image plane. The images are described using a Cartesian angular coordinate system. We will consider axially symmetric lens models, and refer to the radial distance from the fiducial ray through the lens center as $\beta=\sqrt{\beta_1^2 + \beta_2^2}$, and also on the image plane $\theta=\sqrt{\theta_1^2+\theta_2^2}$. The thin lens equation describes the mapping between the unlensed and lensed coordinate systems,
\begin{equation}
\mathbf{\beta} = \mathbf{\theta} - \frac{D_\text{ds}}{D_\text{s}} \mathbf{\hat{\alpha}} = \mathbf{\theta}-\mathbf{\nabla_\theta} \psi,
\end{equation}
where $D_{\rm ds}$ and $D_{\rm s}$ is the angular diameter distance between the lens and the source, and the observer and the source respectively, the reduced deflection angle $\mathbf{\alpha} =
\frac{D_\text{ds}}{D_\text{s}} \mathbf{\hat{\alpha}}$ can be written
in terms of the effective lens potential such that
\begin{equation}
\mathbf{\alpha} = \mathbf{\nabla_\theta} \psi(\theta).
\label{defl}
\end{equation}
We follow the sign convention that deflection angles that bend rays toward the lens (converging lensing) are positive, and deflections away from the lens (diverging lensing) are negative. In this work, we study diverging lenses with negative deflection angle. In general, the magnification produced by a lens is inversely related to the Jacobian $A$ of the lens mapping, such that $\mu^{-1}=\det(A)$ \citep{narayan}. In the case of an axially symmetric lens, the magnification simplifies,
\begin{equation}
\mu^{-1}=\frac{\beta}{\theta} \frac{\text{d} \beta}{\text{d} \theta}.
\end{equation}
The eigenvalues of the Jacobian give the magnifications in the tangential and radial directions,
\begin{equation}
  \mu_\text{t}^{-1} =  \frac{\theta}{\beta},
  \qquad\qquad
  \mu_\text{r}^{-1} =  \frac{\text{d}\theta}{\text{d}\beta}.
\end{equation}
Critical lines arise where $\det(A)=0$, so the tangential and radial critical curves of a lens are found where these expressions vanish. As we show in the following sections, the diverging lenses we are considering produce only radial critical curves. For a given source position, the sum of all the image magnifications gives the total magnification produced by the lens,
\begin{equation}
\mu_\text{T} = \sum_i |\mu_i|.
\end{equation}

The Jacobian can be expressed in terms of the convergence and shear of the lens. The convergence describes the magnification as the increase in the size of the image,
\begin{equation}
\kappa(\theta) = \frac{1}{2}\nabla_\theta^2 \psi(\theta).
\label{conv}
\end{equation}
Image distortion caused by lensing is described by the two components of the shear \citep{narayan},
\begin{equation}
\gamma_1(\theta) = \frac{1}{2}\left( \frac{\partial^2 \psi}{\partial\theta_1^2} - \frac{\partial^2 \psi}{\partial \theta_2^2} \right)
\end{equation}
\begin{equation}
\gamma_2(\theta) = \frac{\partial^2 \psi(\theta)}{\partial\theta_1 \partial\theta_2},
\end{equation}
with $\gamma=\sqrt{\gamma_1^2 + \gamma_2^2}$ the magnitude of the shear. A positive or negative $\gamma$ means a corresponding tangential or radial image stretching.

To demonstrate the resulting image morphology from a particular configuration of lens and source, we use the crosshair and implicit function imaging methods to map points from the source to the image plane \citep{schrammKayser, kayserSchramm}.

%%%%%%%%%%%%%%%%%%%%%%%%%%%%%%%%%%%%%%%%%%

\section{A simple diverging lens model}
\label{sec:divergingLens}

General relativity produces the gravitational lens deflection due to a point mass in analogy with the deflection from an optical medium that has a spatially varying index of refraction \citep{gordon},
\begin{equation}
{\rm n}_{\rm r}(r) \approx 1-\frac{2}{c^2} \Phi(\mathbf{r}) = 1+\frac{2GM}{c^2r},
\label{generalIndex}
\end{equation}
where $\Phi(\mathbf{r})$ is the usual Newtonian potential of the lens and $r$ is the radial distance from the origin \citep{SEF}. For a simple example of diverging lensing, consider the effect of some potential with $\Phi(r)>0$ everywhere. This has an analogous effect on electromagnetic radiation as Rutherford scattering of charged particles from a repulsive potential \citep{selmke2, selmke3, selmke}. In this section we summarize the properties of a simple diverging lens in direct analogy with the point mass lens, but for which gravity is repulsive instead of attractive (i.e., for which mass is negative). The negative-mass lens was first described by \citet{safonova} in a study of exotic lenses, however no connection was made to plasma lensing in that work.

The diverging point lens is described by simply flipping the sign of the effective lens potential of a point mass
\begin{equation}
\psi=- \theta_0^2 \ln{\theta}.
\label{divLensPot}
\end{equation}
For the point mass lens we simply treat $\theta_0$ as constant with respect to $\theta$ and identify it with the ``Einstein radius'' such that $\theta_0 =\theta_E$. In Section\,\ref{sec:power-law} we will work out the details of $\theta_0$ for a more involved scenario that includes the effect of dispersion in the plasma lens model.

The thin lens equation for the negative mass point lens gives two solutions in general,
\begin{equation}
\theta=\frac{1}{2}\left( \beta \pm \sqrt{\beta^2-4\theta_0^2}\right).
\end{equation}
These solutions are purely imaginary for the region
\begin{equation}
\beta \leq 2 \theta_0.
\label{sourceLimit}
\end{equation}
We define this area of the source plane as the ``exclusion region'', where the most significant demagnification of the source occurs. More generally, for circularly symmetric diverging lenses with one or more nested caustics, we define the interior of the innermost caustic as the exclusion region. Sources within the innermost caustic are strongly de-magnified in general. In the case of the negative mass point-lens, no images occur at all for sources in the exclusion region of the lens, as radiation from the source is completely refracted away from an observer. This is due to the diverging nature of the lens, which repels rays from the center.

We write the magnification of the images in terms of the source coordinates, such that
\begin{equation}
\mu_\pm=\frac{1}{2}\left( 1 \pm \frac{\beta^2 - 2\theta_0^2}{\beta \sqrt{\beta^2 - 4\theta_0^2}} \right).
\end{equation}
This differs from the typical converging lens due to the minus signs in the numerator and denominator, which are positive in the usual gravitational case. The total magnification from the diverging lens in the geometrical limit is
\begin{equation}
\mu_{\rm T}=|\mu_+| + |\mu_-|= \frac{\beta^2 - 2\theta_0^2}{\beta \sqrt{\beta^2 - 4\theta_0^2}}
\label{mu_lens_plasma}
\end{equation}
for $\beta>2\theta_0$, and $\mu_{\rm T}=0$ for $\beta\leq2\theta_0$. The minus sign in the denominator implies that the magnification diverges at the edge of the exclusion region.

The radial and tangential magnification of the negative mass diverging lens is opposite to the gravitational point mass lens
\be
\mu_{\rm r,t}={1\over 1\mp \theta_0^2/\theta^2}.
\ee
For the gravitational lens the images are tangentially extended, and may occur on opposite sides of the lens center. Rather than tangentially extended images, the diverging lens produces radially extended images on the same side of the lens center. This behaviour is shown in Figures 4-9 in \citet{safonova}.

While the negative mass point lens is an interesting demonstration of diverging lensing, it is based on an unphysical index of refraction. We have not considered any dispersive phenomena, since we simply take the quantity $\theta_0$ in the lens potential (Eq.\,\ref{divLensPot}), which sets the angular scale of the lens, as a fixed constant. However, in the next section we will describe the effect of a plasma lens by including frequency-dependence in the angular scale $\theta_0$.

\section{A power-law plasma lens}
\label{sec:power-law}

Let us consider the dispersion of electromagnetic rays through a plasma lens. This can be used to describe the propagation of very low-frequency radiation through intervening plasma density inhomogeneities \citep{ESE0, ESE1, ESE2, ESE3}. We neglect the magnetic field along the line of sight and write the
frequency-dependent ``effective potential'' by comparing the index of refraction of cold plasma
\begin{equation}
{\rm n}_{\rm r}^2 = 1 - \frac{\omega_\text{e}^2}{\omega^2 }
\end{equation}
with Eq.\,\ref{generalIndex}, which gives
\begin{equation}
\Phi \approx \frac{c^2 \omega_e^2}{4 \omega^2}
\end{equation}
in the limit of a large observational frequency $\omega \gg \omega_\text{e}$. The plasma frequency
\begin{equation}
\omega_\text{e}^2 = \frac{e^2 n_\text{e}(r)}{\epsilon_0 m_\text{e}}
\end{equation}
depends on the electron number density $n_\text{e}(r)$. The other quantities in this expression are electron charge $e$, the mass of the electron $m_\text{e}$ and permittivity of free space, $\epsilon_0$ \citep{plasmaBook, perlickBook}.

In order to obtain the effective lensing potential, we consider a ray passing through the plasma lens with $\hat{z}$ directed parallel to the fiducial direction and perpendicular to the lens plane. Using the low-frequency limit, and the shorthand
\begin{equation}
N_\text{e}(\theta) = \int n_\text{e} dz
\end{equation}
along with the classical electron radius
\begin{equation}
r_\text{e} = \frac{e^2}{4 \pi \epsilon_0 m_\text{e} c^2 },
\end{equation}
we obtain the effective potential for a plasma lens
\begin{equation}
\psi(\theta) = \frac{D_\text{ds}}{D_\text{s} D_\text{d}} \frac{1}{2\pi} r_\text{e} \lambda^2 N_\text{e}(\theta),
\end{equation}
where $\lambda = 2 \pi c / \omega$ is the wavelength of the ray.
All other lensing properties can be derived from the potential in analogy with gravitational lensing. Additional details on the derivation of the plasma lens potential can be found in \citet{ESE4}.

Let us consider the example of a plasma lens with an electron density that decreases as a function of distance from the center of the plasma distribution, such that
\begin{equation}
n_\text{e}(r)=n_0\frac{R_0^h}{r^h}
\end{equation}
where $h > 0$, and $n_0$ is a constant representing the electron density at a constant characteristic radius $r=R_0$ \citep{review}. Using the impact parameter of a ray on the lens plane $b$, and distance along the line of sight $z$ we express the three-dimensional radial coordinate $r=\sqrt{ b^2 + z^2}$. Then Eq.\,\ref{defl} gives
\begin{equation}
\hat{\alpha}(b) = \frac{\lambda^2 r_\text{e} n_0 R_0^h}{\pi} \int_0^\infty \frac{\partial}{\partial b} \left(  b^2 + z^2 \right)^{-\frac{h}{2}} dz.
\end{equation}
This expression can be evaluated analytically (see \citet{BKT09} and \citet{BKT10} for details), which leads to the reduced deflection angle
\begin{equation}
\alpha(\theta) = - \lambda^2 \frac{D_\text{ds}}{D_\text{s} D_\text{d}^h} \frac{r_\text{e} n_0 R_0^h}{\sqrt{\pi} } \frac{\Gamma\left( \frac{h}{2} + \frac{1}{2}\right)}{\Gamma\left( \frac{h}{2} \right)} \frac{1}{\theta^h}.
\label{plDefl}
\end{equation}
Converting this expression from wavelength to frequency, and using the definition of the classical electron radius, this expression is identical to the deflection angle from the plasma lens contribution in \citet{BKT09}, who considered a more general case, including the effects of both gravitation and plasma.

Let us define the angular scale length for this lens as
\begin{equation}
\theta_0 = \left( \lambda^2 \frac{D_\text{ds} }{D_\text{s} D_\text{d}^h} \frac{r_\text{e} n_0 R_0^h }{\sqrt{\pi}} \frac{\Gamma\left( \frac{h}{2} + \frac{1}{2} \right)}{\Gamma\left( \frac{h}{2} \right)} \right)^\frac{1}{h+1}.
\label{plasmaScale}
\end{equation}
The effective potential for the power-law plasma lens is then simply stated in terms of the scale length as
\begin{equation}
\psi(\theta) = \left\{
\begin{array}{ll}
  \dfrac{\theta_0^{h+1}}{(h-1)} \dfrac{1}{\theta^{h-1}}, & h \neq 1 \\
  \\
- \theta_0^2 \ln \theta, & h=1
\end{array}\right. .
\label{dispPsi}
\end{equation}
In the case that $h=1$, we arrive at the same thin lens equation as the negative point mass lens in Section\,\ref{sec:divergingLens}, with the exception that $\theta_0 \rightarrow \theta_0(\lambda)$. We will return to this point in Section \ref{subsec:exotica}. The analytical solutions for the image positions for $h=2$, $3$ are given in Appendix \ref{appA}.

The main difference introduced by dispersion is that the wavelength-dependence of the scale factor causes the exclusion region of the lens to shrink at short wavelengths, such that the plasma lens is totally transparent in the limit of high-frequency radiation.

The power-law lens provides a simple form for the convergence and the shear
\begin{flalign}
  \kappa = &{h-1 \over 2}{\theta_0^{h+1} \over \theta^{h+1}},\\
  \gamma = &-{h+1 \over 2}{\theta_0^{h+1} \over \theta^{h+1}}.
\end{flalign}
The negative shear value represents one of the main differences between a diverging lens and a converging lens. For a diverging lens the direction of the image distortion is radial. The
magnification can be stated
\begin{equation}
\mu^{-1} = 1+(1-h)\frac{\theta_0^{h+1}}{\theta^{h+1}} - \frac{h \theta_0^{2(h+1)}}{\theta^{2(h+1)}}.
\label{plmag}
\end{equation}
The eigenvalues of the Jacobian for the power-law lens are
\begin{flalign}
\mu^{-1}_{\rm r} &= 1-h\dfrac{\theta_0^{h+1}}{\theta^{h+1}};\\
\mu^{-1}_{\rm t} &= 1+\dfrac{\theta_0^{h+1}}{\theta^{h+1}},
\end{flalign}
which shows the power-law plasma lens has only a radial critical curve, contrasting with the point mass gravitational lens. We can also write the corresponding position of the critical curve and caustic in terms of image and source coordinates,
\begin{flalign}
\theta_\text{critical} &= h^{1\over h+1} \theta_0,\\
\beta_\text{caustic} &= (h+1)\frac{\theta_0}{h^\frac{h}{h+1}}.
\end{flalign}

%%%%%%%%%%%%%%%%%%%%%%%%%%%%%%%%%%%%%%%
\begin{center}
\begin{figure*}
  \includegraphics[bb= 42 220 559 560, clip, scale=0.95]{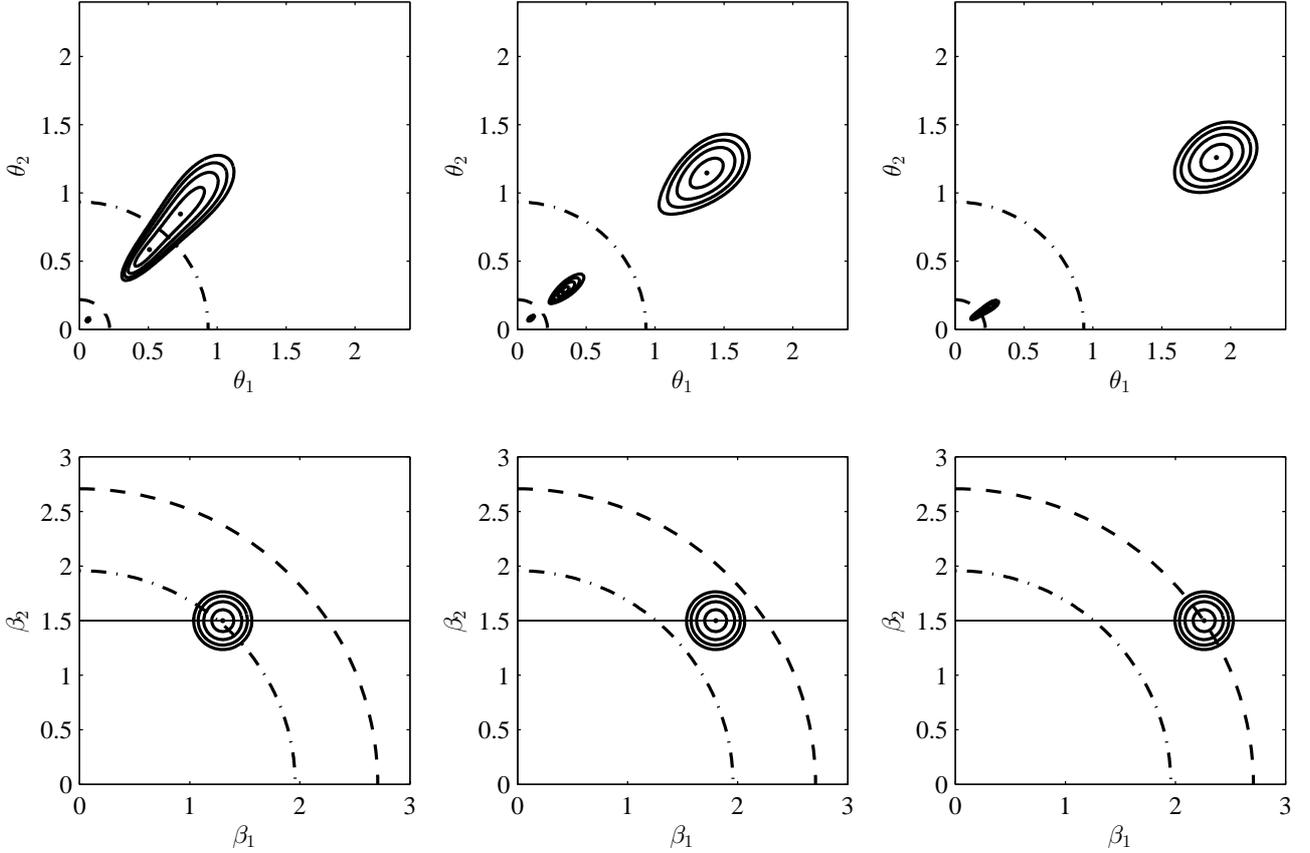}
  \caption{Properties of the softened power-law lens with power-law index $h=1$, characteristic scale angle $\theta_0=1$ (arbitrary units) and core size $\theta_C=0.2$. The source has an impact parameter $\beta_2=1.5$. Bottom row shows the caustics and source position. The radial caustics are plotted as dashed and dashed-dotted curves and the path of the source is the solid black line. The top row shows the resulting critical curves and image configuration. The dashed and dashed-dotted critical curves map to the corresponding caustics. Left: the source crosses from the interior caustic, passing from the exclusion region into the inter-caustic region in which three images are formed at the exterior critical curve. Center: In the inter-caustic region the source forms three lensed images. Right: The source crosses the outer caustic and a pair of images merge at the inner critical curve.}
\label{figPL1}
\end{figure*}
\end{center}
\begin{center}
\begin{figure}
\includegraphics[bb= 150 245 440 545, clip, scale=0.8]{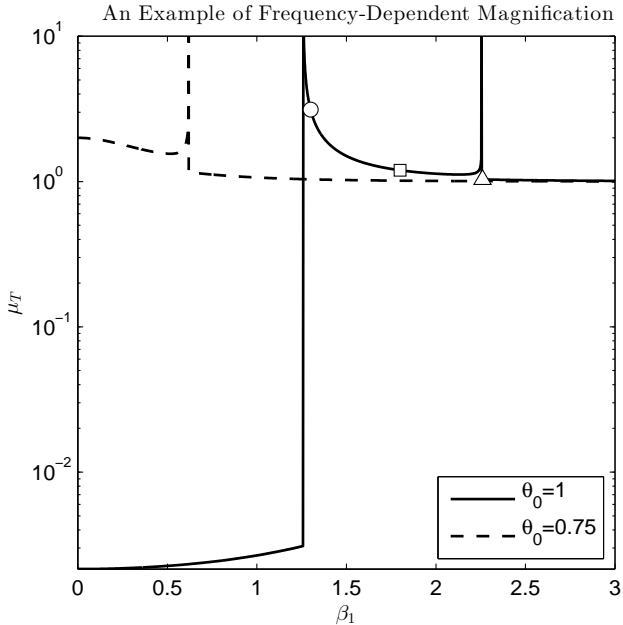}
\caption{ The magnification along the path traversed by the source for two characteristic angular scales, $\theta_0=1$ (solid line) and $\theta_0=0.75$ (dashed line). The y-axis logarithmically scaled to emphasize the low magnification of the image within the exclusion region. The left, center and right positions of the source in Figure \ref{figPL1} are marked with a circle, square and triangle, respectively. The core size is $\theta_C=0.2$ in both cases. The magnification is calculated for a point source with impact parameter $\beta_2=1.5$, and diverges at the position of the caustic crossings.}
\label{figPLMag}
\end{figure}
\end{center}
\subsection{Softened power-law plasma lenses}
\label{subsec:soft}
\begin{figure}
\includegraphics[height=7.5cm,width=8.5cm]{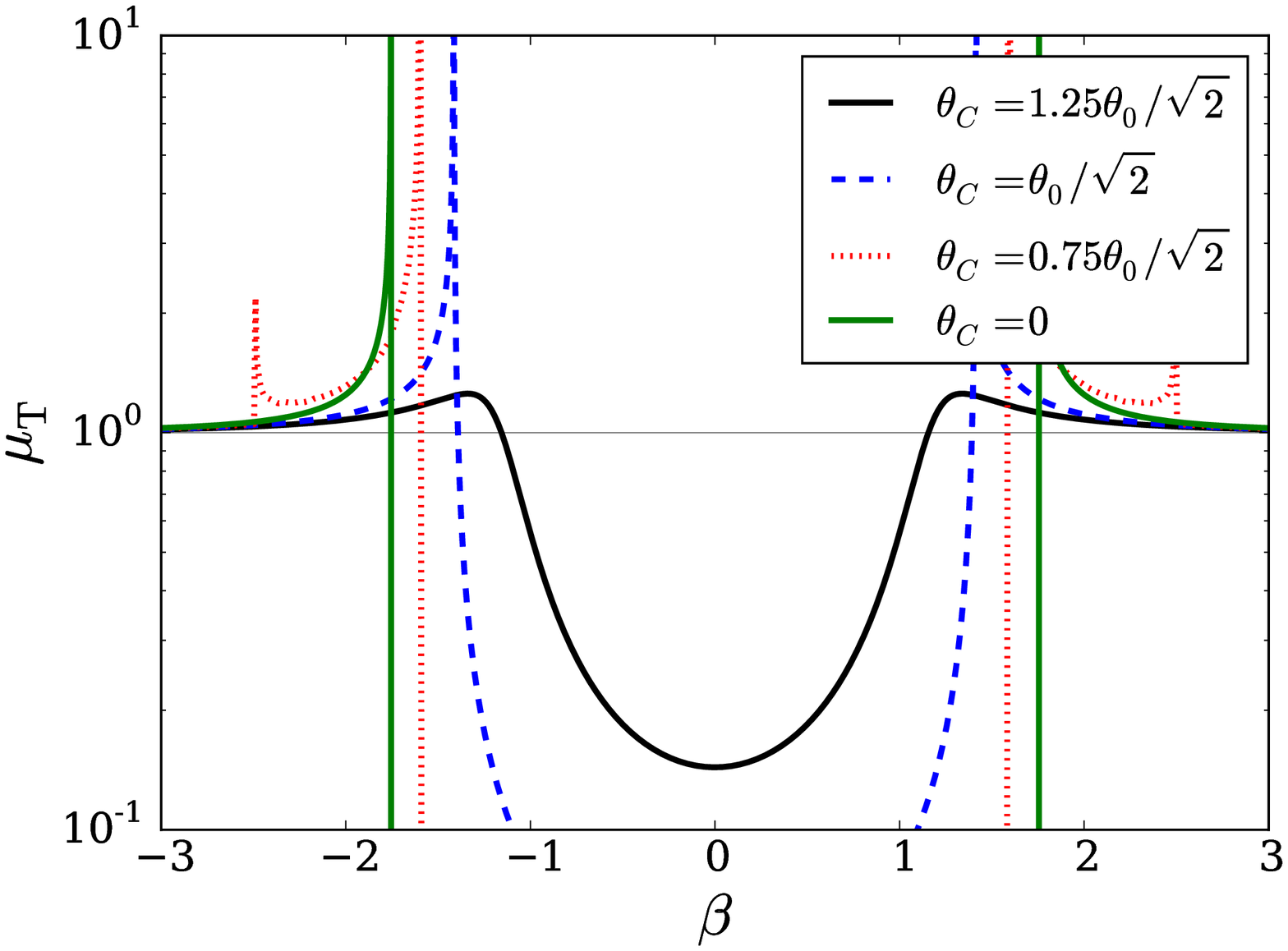}
\caption{Magnification curves produced by the softened power-law lens ($h=3$) from a source passing through the center of the lens. The dotted, dashed and solid curves represent lenses with two, one and no caustics respectively.}
\label{figsofth3}
\end{figure}
%%%
While a power-law lens serves as a useful example to explore the effect of diverging lensing on image formation, it may be considered an unphysical model because of the singularity of the electron density at the center of the lens. A physically realistic lens distribution should not include this feature. In view of this shortcoming, we generalize the power-law model by including a finite core, with angular core radius $\theta_\text{C}$. This is done by simply adding the core size in quadrature in the usual expression for the radius $\theta \rightarrow \sqrt{\theta^2+\theta_\text{C}^2}$. In this case the lens potential becomes
\begin{equation}
\psi(\theta) = \left\{
\begin{array}{ll}
  \dfrac{\theta_{0}^{h+1}}{(h-1)} \dfrac{1}{\left(\theta^2 + \theta_\text{C}^2\right)^\frac{h-1}{2}}, & h \neq 1 \\
\\
- \theta_{0}^2 \ln\left[ \left( \theta^2 + \theta_\text{C}^2 \right)^\frac{1}{2} \right], & h=1
\end{array}\right.
\label{finite_core_potential}
\end{equation}
with the angular scale given in Eq.\,\ref{plasmaScale}. The softened potential gives the deflection angle
\begin{equation}
\alpha(\theta) = - \theta_0^{h+1} \frac{\theta}{\left( \theta^2 + \theta_\text{C}^2\right)^\frac{h+1}{2}},
\end{equation}
and the convergence and shear
\begin{flalign}
  \kappa=&{1\over 2}\eck{-\dfrac{2\theta_0^{h+1}}{(\theta^2+\theta_C^2)^{h+1 \over 2}} + \dfrac{(h+1)\theta_0^{h+1}\theta^2}{(\theta^2+\theta_C^2)^{h+3\over 2}}},\\
  \gamma=&-{1\over 2}\dfrac{(h+1)\theta_0^{h+1}\theta^2}{(\theta^2+\theta_C^2)^{h+3\over 2}}.
\end{flalign}
Again we can see that the shear value is negative. In addition to the inverse magnification,
\begin{equation}
\begin{array}{ll}
\mu^{-1}= & 1+\dfrac{\theta_0^{h+1}}{\left( \theta^2+\theta_\text{C}^2 \right)^\frac{h+1}{2}} \left[2-\dfrac{(h+1)\theta^2}{\theta^2+\theta_\text{C}^2} \right] \\
 & + \frac{\theta_0^{2h+2}}{\left( \theta^2+\theta_\text{C}^2 \right)^{h+1}} \left[ 1-\dfrac{(h+1)\theta^2}{ \theta^2+\theta_\text{C}^2} \right],
\end{array}
\end{equation}
we also have the eigenvalues of the Jacobian, giving the radial and tangential magnifications
\begin{flalign}
\mu^{-1}_{\rm r}&=1+\dfrac{\theta_0^{h+1}}{(\theta^2+\theta_\text{C}^2)^{(h+1)/2}} - \dfrac{(h+1) \theta_0^{h+1}\theta^2 }{(\theta^2+\theta_\text{C}^2)^{(h+3)/2}},\\
\mu^{-1}_{\rm t}&=1+\dfrac{\theta_0^{h+1}}{(\theta^2+\theta_\text{C}^2)^{(h+1)/2}}.
\label{eq:mu-softpower}
\end{flalign}
In the limit $\theta_\text{C} \rightarrow 0$ or $\theta\gg\theta_\text{C}$, these expressions reduce to the power-law lens equations.

However, for finite $\theta_\text{C}$ the behaviour of the softened power-law
lens substantially differs from that of the singular case. By
evaluating $\mu^{-1}_{\rm r}=0$, one can solve for the position of the critical
curve. As we can see, two critical curves may exist depending on the
lens scale $\theta_0$ and core size $\theta_C$
(Fig.\,\ref{figPL1}). A small core around the lens centre causes a second critical curve to
appear near the lens center, interior to the first critical curve. As
$\theta_C$ is increased, the inner critical curve moves outwards and
merges with the outer critical curve at the value of transition, and
then both disappear. Thus, a large core in the lens centre, due to a
smooth distribution of electrons, reduces the overall strength of the
lens. On the other hand, if $\theta_C$ decreases to $0$, the inner
critical curve shrinks to the lens centre and disappears at
$\theta_C=0$, i.e. reduces to the singular case. We analytically solve the magnification for two simple cases ($h=1,3$), and list the conditions for the appearance of
multiple radial critical curves in Table\,\ref{tab:softpow-cc}. The lens profiles
with other values of the power index $h$ have similar behaviours with different
transition values which can be solved numerically. Besides the critical curve morphology, the presence of a finite core also changes the nature of the exclusion region for all non-vanishing values of $\theta_C$. When two critical curves are present, a single image is formed in the exclusion region. However, the magnification of this image remains generally small for sources near the center of the exclusion region. As the source approaches the edge of the exclusion region the magnification increases smoothly. Beyond the critical value of the core size, when no caustics are formed, the magnification of an image near the lens center grows as $\theta_C$ is increased.

In Fig.\,\ref{figPL1} we show an example using an $h=1$ power-law lens with characteristic scale $\theta_0=1$ and core size $\theta_C=0.2$ in arbitrary units. This scenario meets the criteria for two radial critical curves to exist. Then, we choose an impact parameter of $\beta_2=1.5$, and pass the source behind the lens. We plot the resulting images for a source comprised of four concentric circles. Within $\beta=2$, a single point image is found for a source in the exclusion region, with low magnification. As the source traverses the caustics, the corresponding radially-stretched images form and split. In the inter-caustic region, three images exist that are radially extended and colinear with one another as well as the lens center. As the source moves outward through the exterior caustic, the two images interior to the critical curves merge together leaving a single image that moves away from the lens and asymptotically approaches the unlensed source in appearance and position. The magnification corresponding to this path is plotted as a function of the source position, shown as a solid line in Fig.\,\ref{figPLMag}. This plot clearly demonstrates the extreme de-magnification of sources inside the exclusion region, and shows the magnification spikes when the source crosses the caustics. The small magnification for the source within the exclusion region smoothly varies between $\mu_\text{T} \sim 2 \times 10^{-3}$ at the origin to $\mu_\text{T} \sim 3 \times 10^{-3}$ at the edge of the exclusion region for the parameters used in the example. One should also notice, the magnification shown in Fig.\,\ref{figPLMag} is for a point source, and can be different for an extended source.

Due to the dependence of the characteristic angular scale on wavelength, when observing a plasma lens at a shorter wavelength the effect of the lens is diminished. A smaller wavelength $\lambda$ may cause $\theta_0$ to shrink with respect to $\theta_{C}$, and the images that are formed, along with the morphology of the light curve can change dramatically.
To illustrate this, we consider an example similar to the case in Fig.\,\ref{figPL1}. We decrease the characteristic scale to $\theta_0=0.75$ but keep all other physical parameters fixed including the impact parameter. In this case, the source crosses only the outer caustic instead of both, and avoids the exclusion region entirely. The corresponding magnification for the $\theta_0=0.75$ example is shown as a dashed line in Fig.\,\ref{figPLMag}.
Moreover, we present the total magnification $\mu_\text{T}$ for a source behind a softened power-law lens with $h=3$ (Fig.\,\ref{figsofth3}) and vanishing impact parameter. Three cases of the lens configuration are shown: two caustics, one caustic and no caustics. Fig.\,\ref{figsofth3} shows the light curve as a function of time in a realistic observation, under the assumption that the background point source or the lens has a constant projected velocity with respect to the observer. Both examples show that the light curve can vary in morphology substantially when observed at different wavelengths.

\begin{figure}
\includegraphics[height=7.5cm,width=8.5cm]{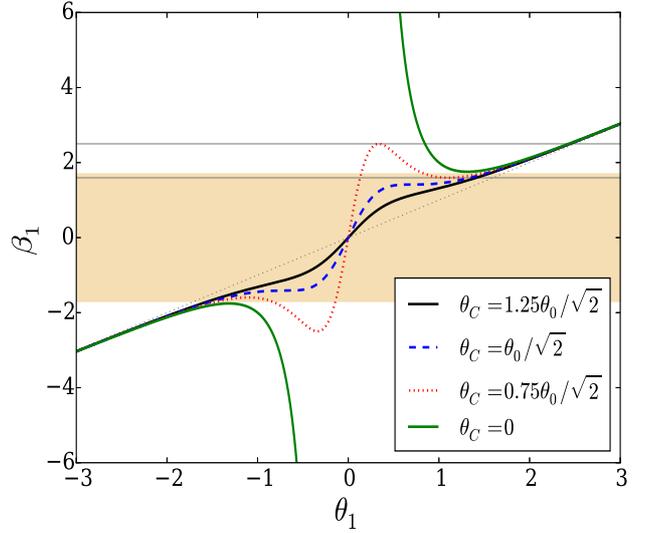}
\caption{Young diagram for the power-law lens with $h=3$ for a source passing through the lens center. The dotted diagonal line shows the unlensed case ($\beta=\theta$). The green line is the singular case, which diverges approaching the origin, and produces no images in the exclusion region (shaded area). The red dotted, blue dashed and black solid curves represent softened power-law lenses with two (super-critical case), one (critical case) and no caustics (sub-critical case) respectively.}
\label{figsofth3Young}
\end{figure}

In Figure \ref{figsofth3Young} we plot a Young diagram, showing the source coordinate $\beta$ as a function of $\theta$ for the softened $h=3$ example. This plot shows the unlensed case as the dotted diagonal line $\beta=\theta$. The singular power-law lens is the bold green line. As we approach the origin from infinity, the singular case diverges. The edge of the exclusion region for this case is defined where the slope of the line vanishes, marked as the shaded region on the figure. The divergence of the curve shows that no images are formed inside the exclusion region. However, the softened power-law is distinct in that no such divergence occurs, and $\beta$ remains continuous as the origin is approached. This means that the softened power-law lens forms an image inside the exclusion region, though as discussed above, this image is generally dim and de-magnified. The super-critical case occurs which the core size is less than the critical value, for example $\theta_C=0.75 \theta_0/\sqrt{2}$. This is shown in Figure \ref{figsofth3Young} as the red dotted line. This curve has two points where the derivative vanishes, marked with horizontal lines, illustrating the dual-caustic structure. Sources between the caustics produce three images, and sources elsewhere produce one. As we approach the origin along this curve, the slope of the line is relatively steep, which shows the de-magnification of the image formed inside the exclusion region. At the critical value of the core size, shown as the dashed blue line, the curve has a plateau and there is only a single caustic formed. In this case, the two caustics have merged. The image formed inside the caustic is brighter than the super-critical case, as can be seen from the shallower slope approaching the origin. As the core size is increased beyond the critical value, a single image is formed for all source positions as no caustics are present. The magnification of an image near the origin grows as $\theta_C$ is increased. However, in all cases with a finite core, the slope is maximal at the origin demonstrating that the most substantial de-magnification occurs within the exclusion region.

\begin{table}
  \begin{tabular}{l|l|l|l}
   model & two crit.curve & one crit.curve  &no crit.curve\\
    \hline
    $h=1$  &$0<\theta_C<\theta_0/2\sqrt{2}$  &$\theta_C=\theta_0/2\sqrt{2}$  &$\theta_C>\theta_0/2\sqrt{2}$\\
    $h=3$ &$0<\theta_C<\theta_0/\sqrt{2}$  &$\theta_C=\theta_0/\sqrt{2}$   &$\theta_C>\theta_0/\sqrt{2}$\\
    \hline
  \end{tabular}
  \caption{The conditions for the softened power-law lens to have different numbers of critical curves. These limits are found by setting the radial inverse magnification to vanish, $\mu_{\rm r}^{-1}=0$, solving for $\theta$ and requiring a single real solution.}
  \label{tab:softpow-cc}
\end{table}

%%%%%%%%%%%%%%%%%%%%

\subsection{Exotic diverging lens equivalency}
\label{subsec:exotica}

We note that the $h=1$ power-law lens, with $\theta_0$ held constant, reproduces the negative mass diverging lens exactly. In fact, there is an equivalent mass $M$ for which a diverging lens will behave identically to the power-law lens with $h=1$ when observed at a wavelength $\lambda_\text{equiv}$. This equivalent wavelength is found by setting the power-law scale constant in Eq.\,\ref{plasmaScale} equal to the Einstein radius $\theta_E$, which gives
\begin{equation}
\lambda_\text{equiv} =  \left( \frac{4 \pi G M}{r_\text{e} n_0 R_0 c^2 } \right)^\frac{1}{2}.
\end{equation}
Thus, for a given plasma density $n_0$ at a distance $R_0$, the $h=1$ dispersive lens behaves identically to a negative-mass lens with mass $M$.

The properties of negative mass gravitational lenses were originally derived by \citet{safonova}. In addition to this unusual model other exotic lenses have been studied in the literature including wormholes \citep{worm2, worm, izumi, asada} and the effect of scalar fields \citep{virb, saf2}. Many of these objects permit diverging gravitational lensing through the violation of one or more energy conditions of general relativity.

One particularly interesting exotic lensing scenario has been studied by \citet{kna} and \citet{bp15}. Rather than the usual Schwarzschild lens, this model supposes a more general form of the metric with the line element
\begin{equation}
ds^2 = A(r) c^2 dt^2 - B(r) dr^2 - r^2 C(r) \left( d \Theta^2 + \sin^2 \Theta d\Psi^2 \right)
\label{lineElement}
\end{equation}
where $\Theta$, $\Psi$ and $r$ are spherical coordinates and $A(r)=1-A_0/r^q$, $B(r) = 1+B_0/r^q$ and $C(r)=1+C_0/r^q$. This metric reproduces the Schwarzschild solution for $q=1$, $A_0=B_0=2GM/c^2$, $C_0=0$. The general form of the metric has a deflection angle given by
\begin{equation}
\hat{\alpha} = \frac{A_0 + B_0}{b^q} \sqrt{\pi} \frac{\Gamma\left( \frac{1+q}{2}\right)}{\Gamma\left( \frac{q}{2} \right)}.
\end{equation}
For $A_0<0$ and $B_0<0$ \citep{kna}, we find the lensing properties exactly align with the power-law plasma lens provided $h \rightarrow q$ and we make the association
\begin{equation}
\lambda_\text{equiv} = \left[ \frac{\pi}{r_\text{e} n_0 R_0^q} \left(|A_0| + |B_0| \right) \right]^\frac{1}{2}.
\end{equation}
Thus, an observation of plasma lensing at a particular wavelength $\lambda_\text{equiv}$ will appear identical to such an exotic lens for given metric components $A_0$, $B_0$ and power-law index $q$. The equivalence is due to the decrease in density of the plasma exactly reproducing the effect of the curvature produced by the exotic metric, such that all the lensing properties are identical to one another.

We point out that it is also possible to find an equivalence between exotic lenses with the other parameters in the power-law lens, such as $n_0$ and $R_0$. However, we focus on $\lambda$ since it is a physical quantity based on observation. At the particular wavelength $\lambda_\text{equiv}$, a power-law lens with given $n_0$, $R_0$ and power-law index $q$ will appear functionally identical to an exotic lens with a given mass $M$. This is especially interesting since the plasma lenses do not require any a priori exotic assumptions to reproduce the exotic lensing behaviour.

%%%%%%%%%%%%%%%%%%%%

\section{The exponential plasma lens}
\label{sec:exponentialLens}
%%%%
\begin{figure}
\centerline{\includegraphics[height=7.5cm,width=8.5cm]{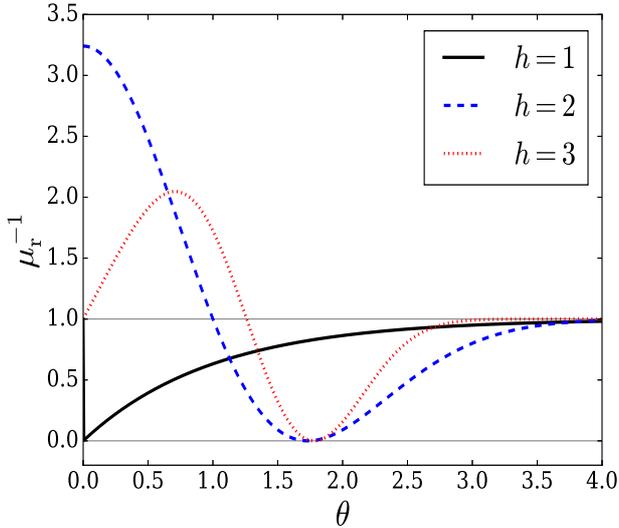}}
\caption{Radial magnification $\mu_{\rm r}^{-1}$ produced by the exponential lens as a function of image coordinate $\theta$. The solid, dashed and dotted curves represent lenses with $h=1$, $h=2$ and $h=3$ respectively.}
\label{figexpon-mucc}
\end{figure}
The most widely used example of a plasma lens to describe ESEs is the Gaussian lens introduced by \citet{cleggFey} to model observations of the extragalactic sources 0954+654 and 1741-038. Since then the model has been the first choice for modeling ESE events. However, the Gaussian is a particular realization of a larger class of model.

We classify the family of exponential lenses in terms of the exponent $h$ by which the radial coordinate distance is raised. A version of the $h=1$ exponential lens was used by \citet{erMao} as a model for a galactic electron distribution. Here we make use of the same basic form but on a smaller scale, describing a local density enhancement for ESEs.
The Gaussian lens of \citet{cleggFey} is an exponential lens with $h=2$. We wish to classify the behaviour of these lenses generally and study their caustic structure. In the derivation \citet{cleggFey} specify the projected electron distribution on the lens plane $N_\text{e}(\theta)$, rather than the full three-dimensional electron distribution $n_\text{e}(r)$ as in the power-law lens \citep{BKT09}. We follow this convention to compare to previous results.

To describe a general exponential lens, we adopt a form for the electron column density in the lens plane,
\begin{equation}
  N_\text{e}(\theta)= N_0 \exp \left( -\frac{\theta^h}{h\sigma^h} \right)
  \qquad (\theta>0),
\end{equation}
with $N_0$ the maximum electron column density within the lens and $\sigma$ as the width of the lens for $h>0$. This particular normalization is chosen to provide an effective lens potential similar to the expression in \citet{newESE2017} for a Gaussian lens ($h=2$). With this lens plane density profile, we evaluate the effective lens potential,
\begin{equation}
\psi = \theta_0^2 \exp \left( - \frac{\theta^h}{h \sigma^h} \right)
\end{equation}
with the angular scale
\begin{equation}
\theta_0=\lambda \left( \frac{D_\text{ds}}{D_\text{s} D_\text{d}}\frac{1}{2\pi} r_\text{e} N_0 \right)^\frac{1}{2}.
\end{equation}
The exponential lens potential leads to the deflection angle
\begin{equation}
\alpha(\theta) = -\theta_0^2 \frac{\theta^{(h-1)} }{\sigma^h} \exp\left( - \frac{\theta^h}{h \sigma^h} \right),
\end{equation}
with the convergence and shear
\begin{flalign}
  \kappa=&\rund{\dfrac{\theta^h}{\sigma^h}-h}\dfrac{\theta_0^2\theta^{h-2}}{2\sigma^h} e^{-\frac{\theta^h}{h\sigma^h}},\\
  \gamma=&\rund{h-2-\dfrac{\theta^h}{\sigma^h}}\dfrac{\theta_0^2\theta^{h-2}}{2\sigma^h} e^{-\frac{\theta^h}{h\sigma^h}},
\end{flalign}
and the inverse magnification
\begin{equation}
\begin{array}{ll}
\mu^{-1} = & 1 + \dfrac{h\theta_0^2}{\sigma^h} \theta^{h-2} \left( 1-\dfrac{\theta^h}{h \sigma^h} \right) e^{ -\frac{\theta^h}{h\sigma^h}} \\
   &  + \dfrac{\theta_0^4}{\sigma^{2h}} \theta^{2(h-2)} \left( h - 1 - \dfrac{\theta^h}{\sigma^h} \right) e^{ - 2 \frac{\theta^h}{h \sigma^h}}.
\end{array}
\end{equation}
Just as the power-law lens, the tangential magnification is seen to be everywhere positive, and the critical curves of exponential lens are caused by the radial magnification
\begin{flalign}
  \mu_{\rm t}^{-1}& =1+\dfrac{\theta_0^2\theta^{h-2}}{\sigma^h} e^{-\dfrac{\theta^h}{h\sigma^h}},\\
  \mu_{\rm r}^{-1}& =1+\dfrac{\theta_0^2\theta^{h-2}}{\sigma^h} \eck{h-1-\dfrac{\theta^h}{\sigma^h}}e^{-\dfrac{\theta^h}{h\sigma^h}}.
  \elabel{expon-murt}
\end{flalign}
The number of critical curves depends on $\theta_0$, $h$ and $\sigma$. Here we discuss the lens of three different exponents ($h=1,2,3$), which reflect the general properties of the exponential lens.

\begin{figure}
\includegraphics[height=7.5cm,width=8.5cm]{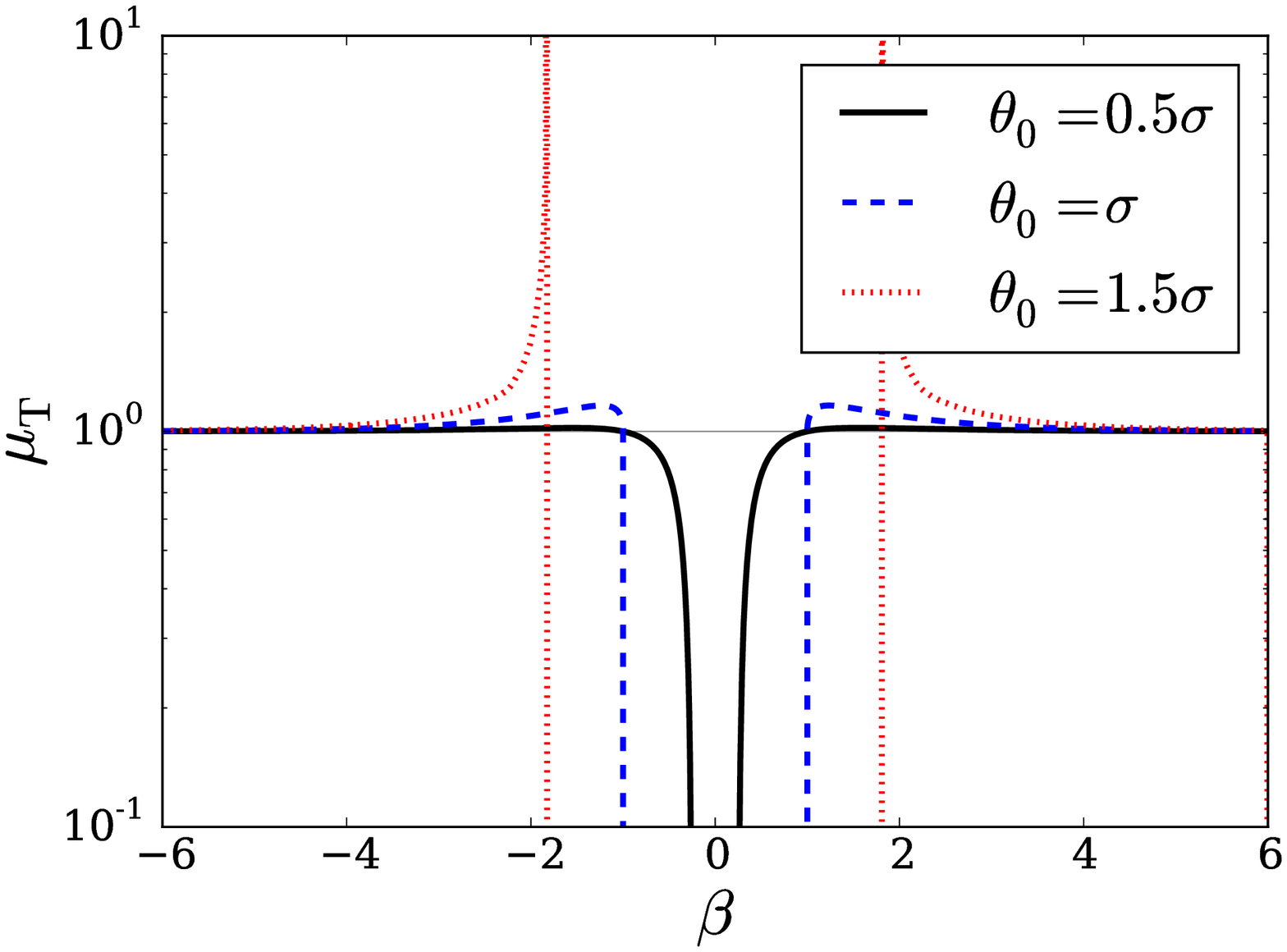}
\includegraphics[height=7.5cm,width=8.5cm]{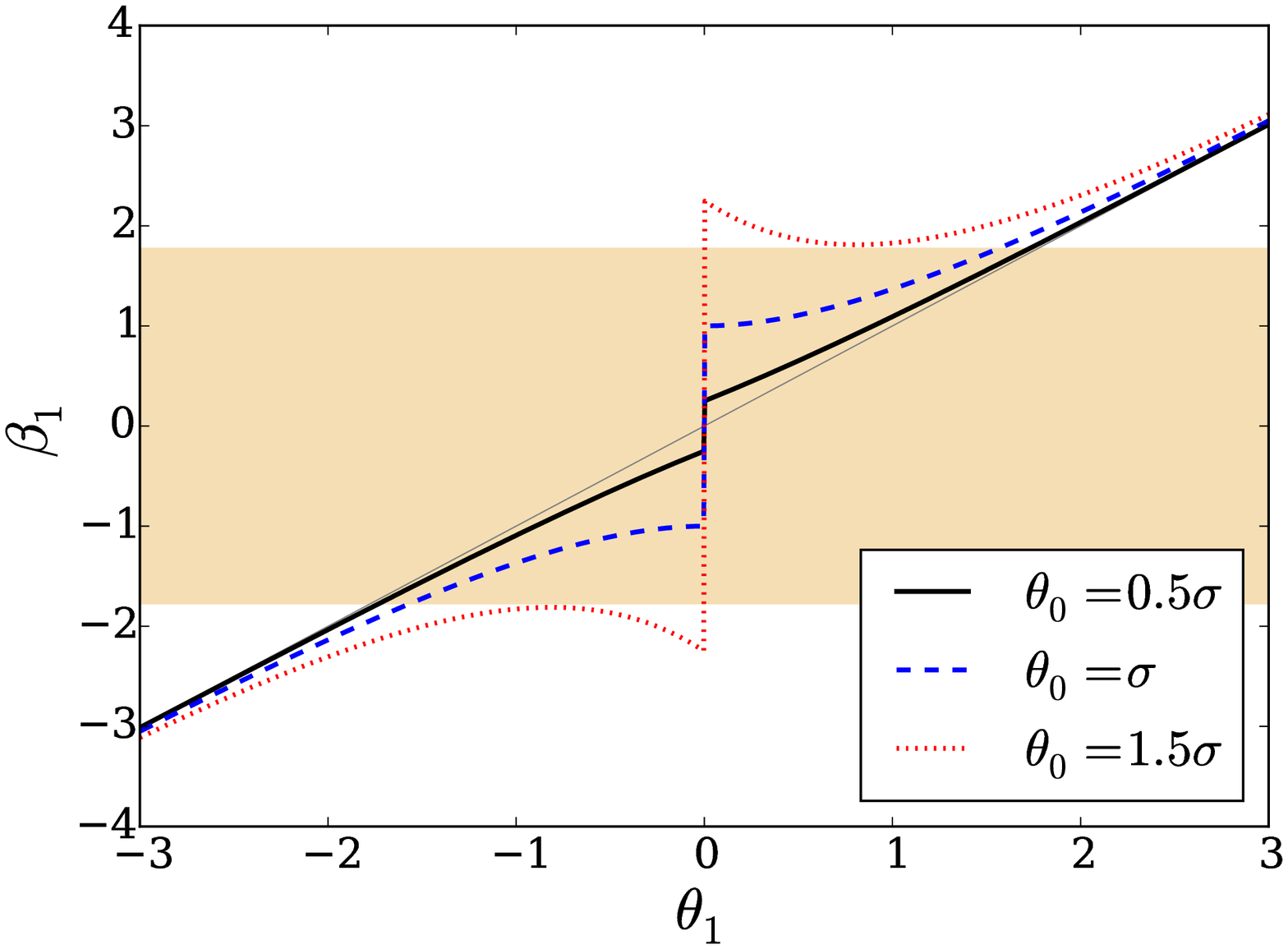}
\caption{Top panel: Magnification curves produced by the exponential lens with $h=1$ from a source passing through the center of the lens. The source coordinate is given in units of $\sigma$. The solid, dashed and dotted lines are the results for $\theta_0 = 0.5 \sigma$, $\theta_0 = \sigma$ and $\theta_0 = 1.5 \sigma$ respectively. Bottom panel: The Young diagram for the $h=1$ exponential lens for a source passing through the lens center. The sub-critical lens behaviour is plotted as a black line, the critical lens is the blue dashed line, and the super-critical behaviour is the red dotted line. Note the presence of the caustic in the super-critical case.}
\label{figExponential_h1}
\end{figure}

In order to study the behaviour of the exponential lens when $h=1$, we plot the Young diagram shown in the bottom panel of Fig. \ref{figExponential_h1}. Let us first consider the sub-critical case, when $\theta_0<\sigma$, shown in the plot as a black line. The $\beta(\theta)$ curve is always increasing away from the origin, so one image is produced that tends to the unlensed case far from the lens. However, as the origin is approached $\theta \rightarrow 0$, the deflection angle tends to a constant value, which defines the extent of the exclusion region. Sources within this region do not produce any images. So the sub-critical exponential lens produces one or no images. When the lens is exactly critical $\theta_0=\sigma$, and the edge of the exclusion region is defined by $\beta=\sigma$ from the thin lens equation. This case also produces only a single image everywhere outside the exclusion region. In this case, as the image appears out of the exclusion region it does so with unit magnification $\mu_\text{T}=1$.

For the super-critical lens, $\theta_0>\sigma$, the Young diagram shows that there is a radial caustic, which corresponds to the edge of the exclusion region. The caustic and corresponding critical curve are given analytically from the radial magnification (Eq.\,\ref{eq:expon-murt}),
\begin{flalign}
  \theta_\text{critical}&=2\sigma{\rm ln}(\theta_0/\sigma), \nonumber\\
  \beta_\text{caustic}&=2\sigma{\rm ln}(\theta_0/\sigma)+\sigma.
  \label{expon1cc}
\end{flalign}
For $\theta_0>\sigma$, zero, one or two images can be produced, depending on the position of the source. As the source moves outward from the lens center and leaves the exclusion region, two images form at the corresponding critical curve and travel in opposite directions. The inward travelling image moves toward the center of the lens, becoming increasingly thin in the tangential direction and vanishes as it approaches the lens center. The image exterior to the critical curve approaches the unlensed image as it travels away from the lens center.

\begin{center}
\begin{figure}
\includegraphics[height=7.5cm,width=8.5cm]{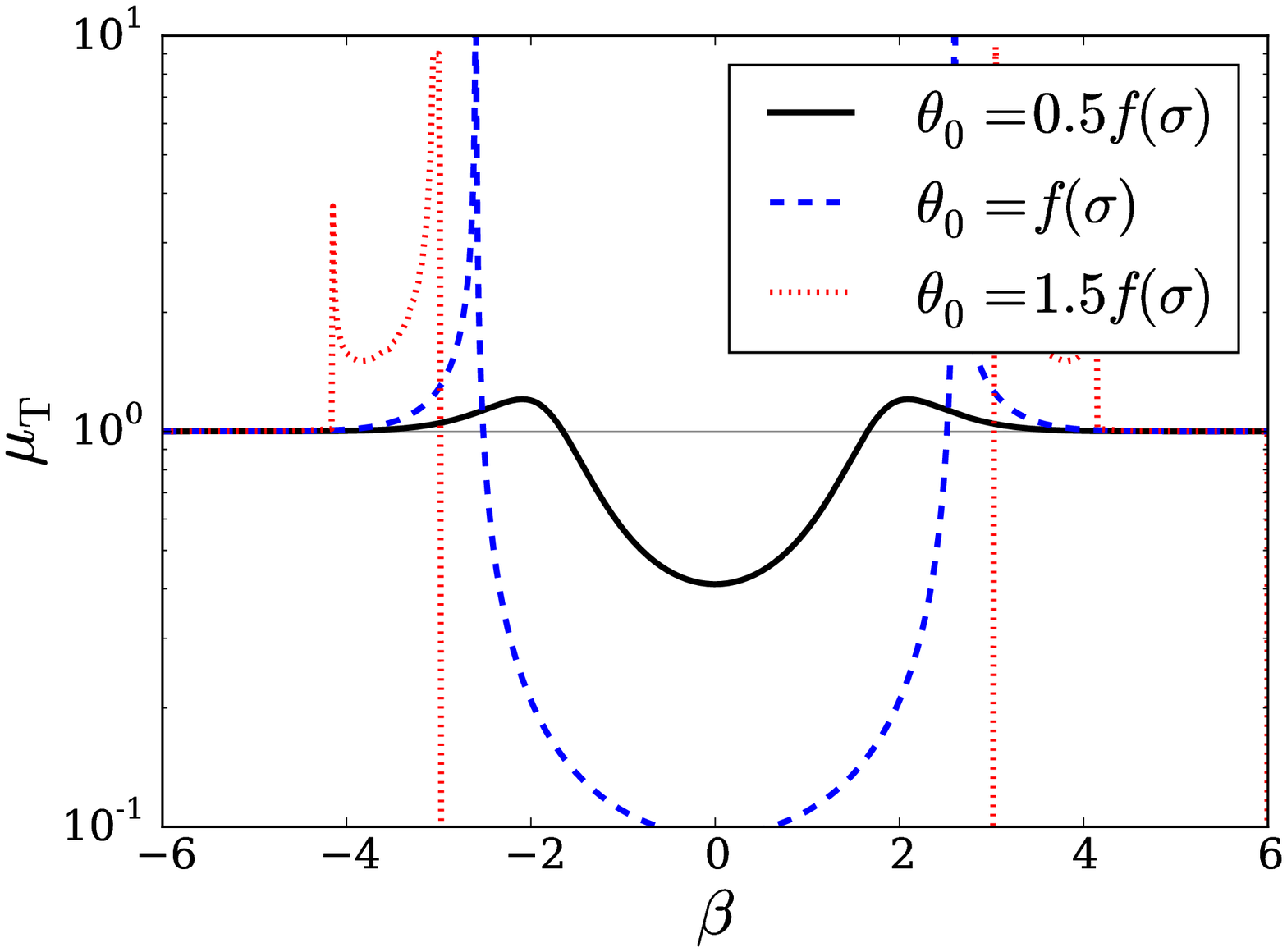}
\includegraphics[height=7.5cm,width=8.5cm]{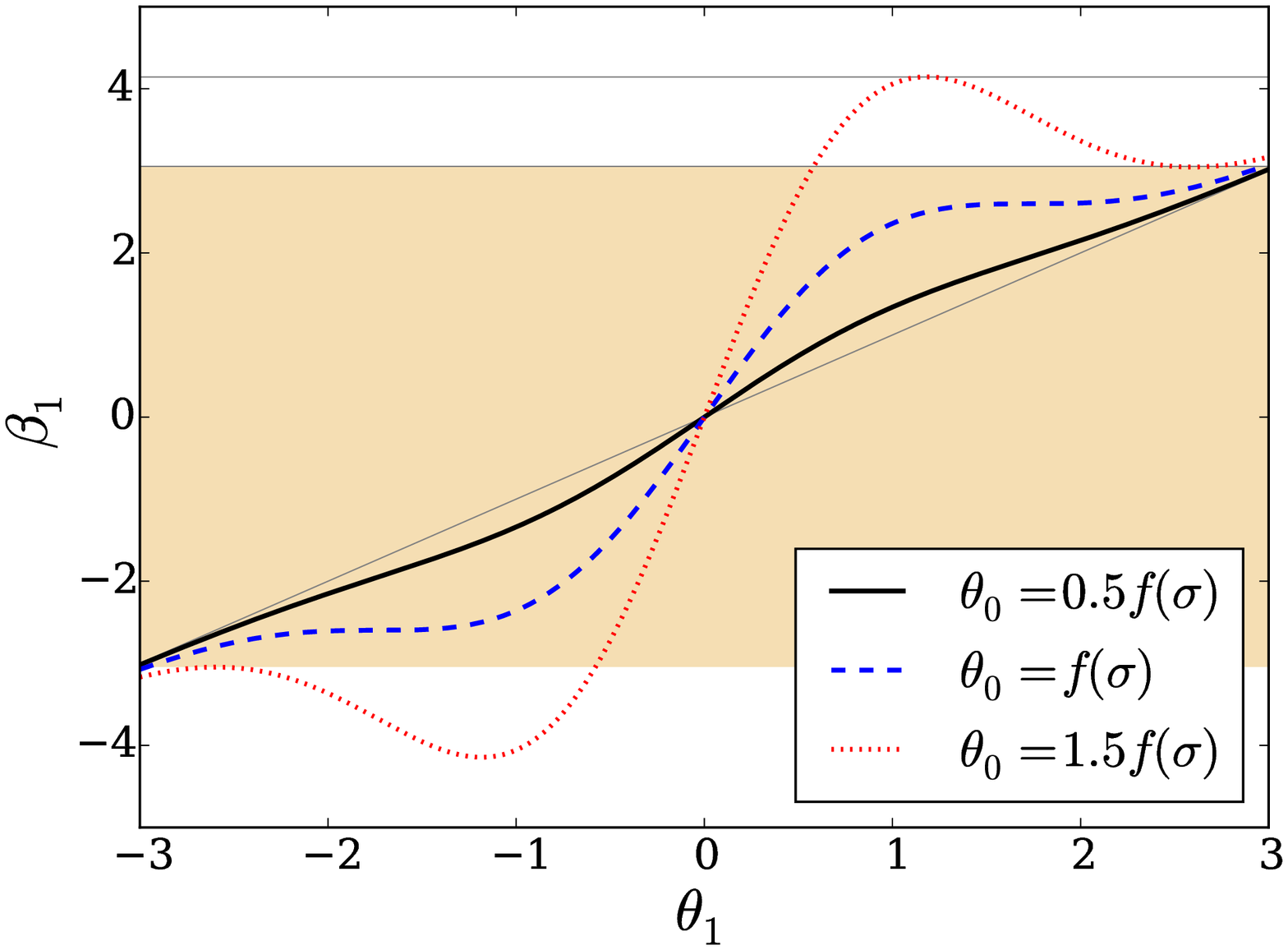}
\caption{Top panel: Magnification curves produced by the Gaussian lens ($h=2$) from a source passing through the center of the lens. The solid, dashed and dotted curves represent lenses with $\theta_0=0.5f(\sigma)$, $\theta_0=f(\sigma)$ and $\theta_0=1.5f(\sigma)$ respectively. The function is defined as $f(\sigma)=\sqrt{e^{3/2} \over 2}\,\sigma$. Bottom panel: The Young diagram for the Gaussian lens. The sub-critical lens behaviour is plotted as a black line, the critical lens is the blue dashed line, and the super-critical behaviour is the red dotted line. Note the presence of the two caustics in the super-critical case.}
\label{figExponential-h2}
\end{figure}
\end{center}

The exponential profile with $h=2$ is a Gaussian lens, for which the
effective lens potential and deflection angle follow the results in
\citet{cleggFey}. Here we only simply summarize its magnification
properties. From the dashed curve in Fig.\,\ref{figexpon-mucc}, we can
see that $\mu_{\rm r}^{-1}(0)>0$ and $\mu_{\rm r}^{-1}(\infty)\rightarrow
1$. Thus, two critical curves will exist if the minimum value of
$\mu_{\rm r}^{-1}$ is smaller than $0$. The analytical relation is given
between $\theta_0$ and $\sigma$.  For small values of the lens angular
scale, $\theta_0 < \sqrt{e^{3/2}/2}\,\sigma$, the lens does not
have a substantial effect. For this configuration, a single image is
produced, and there is no critical curve or exclusion region.  As
$\theta_0=\sqrt{e^{3/2}/2}\,\sigma$, both the critical curve and
exclusion region emerge. When $\theta_0 > \sqrt{e^{3/2}/2}\,\sigma$, a
second critical curve appears, and three lensed images can be
produced. In this configuration the lens leads to a large exclusion
region and two sets of caustic spikes on each side of the lens center,
similar to the softened power-law lens. With the increase of the lens
angular scale $\theta_0$ the second critical curve moves outwards from
the origin. These cases are shown in the top panel of Fig.\,\ref{figExponential-h2},
 and the Young diagram for the Gaussian lens is in the lower panel of the figure.

\begin{center}
\begin{figure}
\includegraphics[height=7.5cm,width=8.5cm]{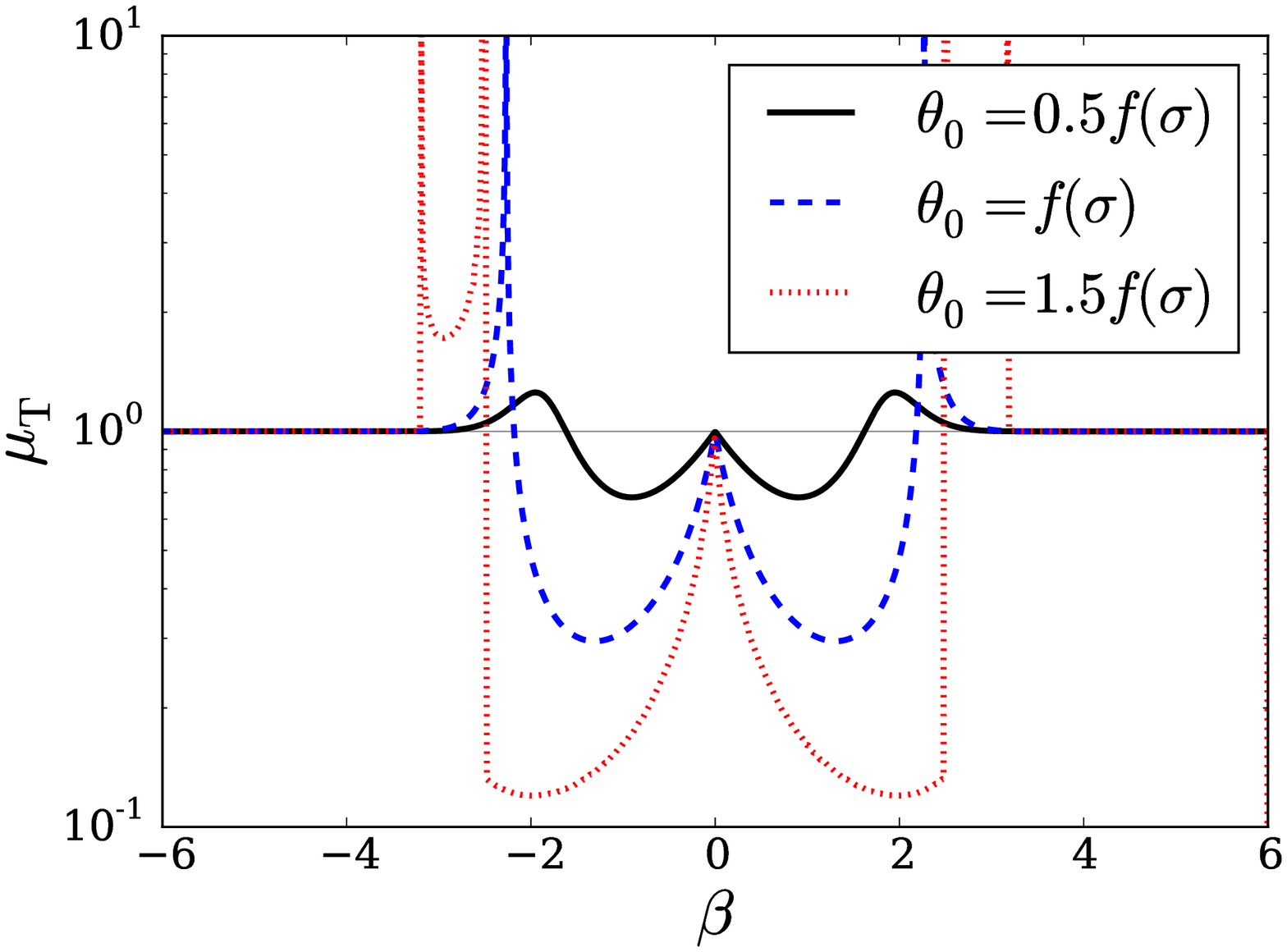}
\includegraphics[height=7.5cm,width=8.5cm]{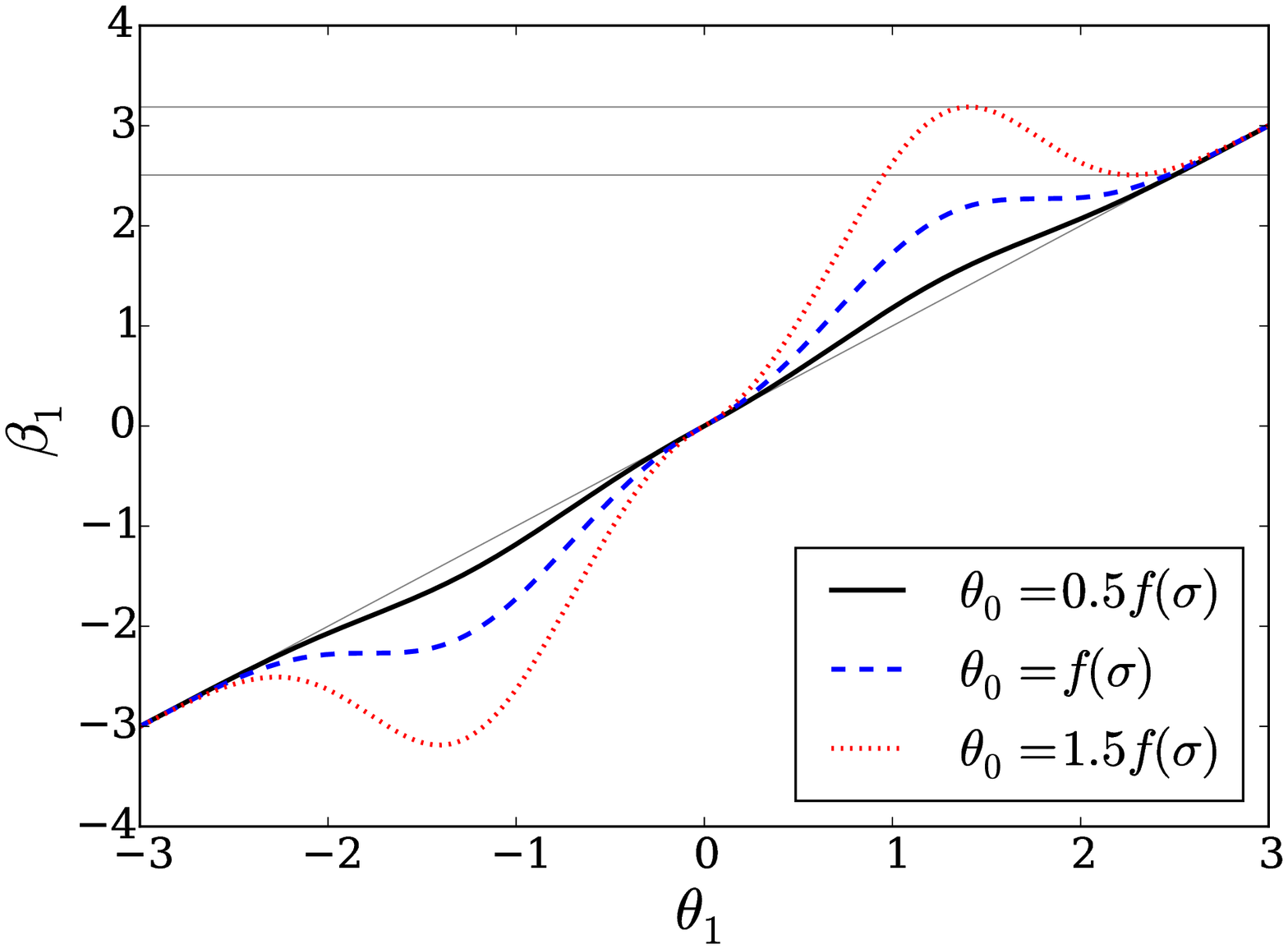}
\caption{Top panel: Magnification curves produced by the exponential lens ($h=3$) from a source passing through the center of the lens. The solid, dashed and dotted curves represent lenses with $\theta_0=0.5f(\sigma)$, $\theta_0=f(\sigma)$ and $\theta_0=1.5f(\sigma)$ respectively. See Eq.\,\ref{eq:exponh3-sigma} for the definition of $f(\sigma)$. Bottom panel: The Young diagram for the $h=3$ exponential lens. The same line styles are used as in Fig.\,\ref{figExponential-h2}.}
\label{figExponential-h3}
\end{figure}
\end{center}

The magnification behaviours for the $h=3$ exponential lens are displayed in the top panel of Fig.\,\ref{figExponential-h3}. In this case, since the deflection angle is proportional to $\theta^2$, it decreases rapidly at small $\theta$, i.e. the lens effects become less significant. Therefore, we can see a peak at $\theta=0$ with $\mu_\text{T}=1$ for all values of $\theta_0$. Thus, within the exclusion region we have a local maximum centered on the origin with unit magnification.
Besides this novel difference, the $h=3$ exponential lens generates similar behaviours as the Gaussian lens.
When $\theta_0$ increases, the dual caustic structure appears once the lens angular scale becomes larger than the transition value
\be
f_{h3}(\sigma)=\dfrac{\sigma}{\sqrt{(\sqrt{7}+1)(3+\sqrt{7})^{1/3}\,{\rm exp}\eck{-\frac{3+\sqrt{7}}{3}}}}.
\elabel{exponh3-sigma}
\ee
We plot the Young diagram for the $h=3$ case in the bottom panel of Fig.\,\ref{figExponential-h3}. The sub-critical case, critical and super-critical cases are plotted as the black, blue dashed and red dotted curves respectively. The two caustics are given for the supercritical case as the horizontal lines. In this plot, the peak in the exclusion region occurs because each of the curves becomes tangential to the unlensed case (diagonal dotted line) for $\beta=0$, and therefore also have unit magnification. For the other exponential profiles with $h>3$, we expect similar magnification behaviours as well, e.g. a central peak with $\mu_\text{T}=1$ and critical curves appearing for sufficiently large $\theta_0$. The transition values for higher $h$ can be solved numerically.

%%%%%%%%%%%%%%%%%%%%%%%%%%%%%%%%%%%%%%%%%%%%%%%%%%%%%%%%%%%%%%%%%%

\section{Discussion}
\label{sec:discussion}

The Gaussian lens (i.e., the exponential lens with $h=2$) has been the most widely used model in describing plasma lensing due to the frequency-dependent magnification produced by the model, which is especially useful in the description of ESEs. It has a distinct caustic structure, with a set of dual radial caustics flanking the origin. We have shown that other members of the exponential lens family can produce equally interesting behaviour. In addition, the softened power-law models can also reproduce the dual caustic structure of the Gaussian lens, an example of which is shown in Fig.\,\ref{figsofth3}.

In this study, we have restricted our analysis of diverging lens models to the geometric limit. In general this is a good approximation for the gravitational lensing of astronomical objects, however diffractive effects can become important for plasma lenses when the angular size of the source is smaller than the Fresnel scale of the lens. Moreover, we address only spherically symmetric lenses in this work, but this does not need to be the case in general, and has been the subject of model development including corrugated sheets and scattering screens for pulsar scintillation \citep{ESE2}, based largely on the asymmetrical Gaussian lens profile. In addition to diverging lensing \citet{lens1} suggested that localized underdense plasma regions will act like converging convex lenses in analogy with gravitational lensing. This fascinating suggestion opens additional avenues for model building. While converging lensing from plasma underdensities has been excluded for particular events \citep{ESE1, ESE3}, this does not conclusively exclude the possibility for all future observations. In addition to parametric diverging lens models, \citet{ESE4} have derived the properties of a plasma lens from inversion of the observed data of PKS 1939-315 in the range $4.2$--$10.8$ GHz using a non-parametric method. This free-form type of model is maximally flexible and can reveal necessary features not considered in the more limited parametric form of a conventional lens model.

In addition to their application in ESEs, plasma lens models have been discussed in regards to the nature of fast radio bursts \citep[FRBs;][]{FRBplasma1}. These events are extragalactic in nature \citep{extragalacticFRB} and one source, FRB121102, has been observed to repeat \citep{repeatingFRB}. Provided that FRBs originate from magnetar activity \citep{magnetarFRB}, compact structures in the host galaxy and the environment around these objects may act as plasma lenses that both amplify and suppress burst flux intermittently given the lens and observer geometry. The plasma structures in the host galaxies of FRBs are expected to be similar to those of our own galaxy. Moreover, the highly structured ejecta in the Crab nebula provide an example that sheets, filaments and dense knots of material can act as lenses to refract radio frequencies \citep{radioScatter}. While FRB events appear to be qualitatively consistent with the properties of Gaussian plasma lenses \citep{FRBplasma1}, the families of models studied here may also be useful for these applications due to the general morphological similarities of the caustics produced by the exponential and softened power-law plasma lenses. Thus, identifying additional lens models that are compatible with the observations is useful for the study of the lensing mechanism proposed to be responsible for these bursts.

In Section \ref{subsec:exotica} we discussed exotic lens models studied by \citet{safonova} that produce diverging lens behaviour. Another model with related properties is the void lens \citep{voidLens1}. This model describes lenses of cosmological scales and while it is unrelated to plasma lensing directly, it provides another example of the dual-caustic structure that is produced by the softened power-law and exponential lens families. The void lens model describes an underdense inhomogeneity in an otherwise isotropic cosmological background that acts as a diverging lens. The spherical volume of material excavated from the background is included in an over-dense shell that limits the range of the potential \citep{embeddedLens,voidLens1}. This phenomena has been used to study lensing effects on the passage of CMB radiation through a swiss-cheese type void \citep{1999MNRAS.309..465A,2013PhRvL.110b1302B,CMBLens, voidLens1}. Besides the diverging behaviour of electromagnetic radiation, voids also behave as diverging lenses for gravitational waves \citep{gw1, gw2, gw3} which are diffracted by mass distributions \citep{takahashi2, starClusterLens, takahashi3}. Our interest is with the general form of the compensated void lens model described by \citet{voidLens1}. The piece-wise lens potential is an example unlike any other studied in this work that produces diverging lens behaviour. Outside of the cosmological setting these types of lensing potentials may be useful on small scales. Since the lensing properties of underdense plasma distributions have been studied previously by \citet{lens1}, adapting a compensated void model on a smaller scale may be useful for describing hollow, bubble-like regions excised from an otherwise uniform distribution of plasma. This may provide another physical model that can produce dual-caustic magnification curves similar to the Gaussian and power-law examples studied in this work.

%%%%%%%%%%%%%%%%%%%%%%%%%%%

\section{Conclusions}
\label{sec:conclusion}

We have studied the behaviour and image formation properties of diverging lens models that are useful in describing ESEs affecting background radio sources and may also provide useful models for the strong refractive events seen in some pulsars. The thin lens formalism has been used in the geometric optics limit to describe the interaction of electromagnetic radiation with plasma in the interstellar medium \citep{cleggFey, ESE2, ESE4}.
We start from a simple toy model refractive lens by simply reversing the effect of gravitation. Despite this simplistic analogy, such a lens identically describes a more realistic dispersive lens at a particular wavelength. The negative-mass point lens produces substantially different lensing effects from the gravitational point lens, such as radial image distortion and strong demagnification near the center of the lens. Instead of an Einstein radius the diverging lens has an exclusion region, where no image can form. From this foundation, we studied two parameterized families of models: the power-law lenses and the exponential lenses. A finite core in the power-law lenses causes dramatic changes in the image properties and the magnification curve. Two radial critical curves can emerge if the core size is smaller than the characteristic scale. For exponential lenses with different indices $h$, the magnification curves have distinct properties. The well-known Gaussian form ($h=2$) produces similar magnification as the softened power-law lens. The exponential lens with $h=1$ can generate only one critical curve and a complete exclusion region, while the lenses with higher index ($h\geqslant3$) generate similar magnification curves except with a bright image at the center of the lens.

Moreover, in both families of lens models, the magnification properties depend on the relation between the characteristic angular scale length $\theta_0$ and other physical parameters. Observations at different wavelengths respond differently to the electron distribution in the lens, and therefore experience a different scale length, which may dramatically change the properties of the magnification curve and the number of critical/caustic curves the lens produces. Therefore, the magnification of radio sources at multiple observation frequencies may appear very different from one another. Thus, these frequency-dependent light curves also improve the physical constraints on models of the plasma density.

\section*{Acknowledgements}
We thank our referee, Artem Tuntsov, whose constructive and valuable comments increased the scope of our work significantly, helped to streamline the flow of the text and increase the overall quality of our manuscript. We thank Oleg Yu. Tsupko and Valerio Bozza for interesting discussion during the preparing of this work. X.E. is funded by Italian Space Agency (ASI) through contract Euclid -IC (I/031/10/0) and acknowledge financial contribution from the agreement ASI/INAF/I/023/12/0. X.E. is also partly support by NSFC Grant No. 11473032. A.R. acknowledges and thanks Samar Safi-Harb for support through the Natural Sciences and Engineering Research Council of Canada (NSERC) Canada Research Chairs Program.

\appendix
\section{Power-law lens solutions}
\label{appA}

In this appendix we discuss how to analytically solve the thin lens equation for the positions of the lensed images using the power-law model for $h=2$ and $h=3$ (Section\,4). The solution for other lens models can be found numerically.

For $h=2$, the thin lens equation becomes a cubic polynomial in the image position,
\begin{equation}
\theta^3- \beta \theta^2 +\theta_0^3 =0.
\label{polyh2}
\end{equation}
The discriminant of this equation is
\begin{equation}
\Delta = -\theta_0^3 \left( 27 \theta_0^3 - 4 \beta^3 \right),
\end{equation}
and determines the number of real solutions of the lens equation. For a given source position $\beta$, one real solution exists when $\Delta<0$, and three solutions exist when $\Delta>0$. At the boundary $\Delta=0$ all three solutions correspond and are a multiple root. This separates the source plane cleanly into two regions, split by the caustic at
\begin{equation}
\beta_\text{c2} = 3 \frac{\theta_0} {4^\frac{1}{3}}
\label{caustic2}
\end{equation}
which is the edge of the exclusion region and introduces the frequency-dependence to the caustic structure of the lens. We can additionally solve for the positions of the images analytically. Reducing the polynomial in Eq.\,\ref{polyh2} to a depressed cubic by means of the substitution
\begin{equation}
\theta = z_k + \frac{\beta}{3}
\end{equation}
gives the solutions for $\Delta>0$ as
\begin{equation}
z_k = \frac{2}{3} \beta \cos \left( \frac{1}{3} \cos^{-1} \left[ 1- \frac{27\theta_0^3}{2\beta^3} \right] -\frac{2\pi k}{3} \right)
\end{equation}
with $k=0,1,2$. However, not all of these real solutions are physical. For example, the $k=2$ solution gives $\theta<0$ and must be discarded since $\theta$ represents the radial distance in the image coordinates. Within the exclusion region, one real solution exists, but this is negative and must also be discarded. Therefore the magnification for $h=2$ is very similar to the $h=1$ case, including vanishing magnification within the exclusion region of the lens.

For $h=3$ the lens equation is a quartic equation,
\begin{equation}
\theta^4 - \beta \theta^3 + \theta_0^4 = 0.
\end{equation}
The limit of the exclusion region is found from the discriminant and gives the radius of the caustic,
\begin{equation}
\beta_\text{c3}=4 \frac{\theta_0}{27^\frac{1}{4}}
\label{caustic3}
\end{equation}
For sources outside of this caustic, the quartic polynomial is solved analytically giving two positive, real solutions,
\begin{equation}
\theta_{\pm} = \frac{\beta}{4} +S \pm \frac{1}{2} \sqrt{ -4S^2 +\frac{3\beta^2}{4} +\frac{\beta^3}{8S} }
\end{equation}
defining
\begin{equation}
S=\frac{1}{2} \sqrt{\frac{\beta^2}{4} + \frac{1}{3} \sqrt{Q+\frac{\Delta_0}{Q}} }
\end{equation}
and
\begin{equation}
Q=\left( \frac{\Delta_1 + \sqrt{\Delta_1^2 -4\Delta_0^3 } }{2} \right)^\frac{1}{3}
\end{equation}
with $\Delta_0 = 12\theta_0^4$ and $\Delta_1 = 27 \beta^2 \theta_0^4$. The remaining solutions for the image positions give negative radial positions and must be discarded since they are unphysical.


\begin{thebibliography}{99}

\bibitem[{{Abbott} {et~al.}(2016{\natexlab{a}}){Abbott}, {Abbott}, {Abbott}, {Abernathy}, {Acernese}, {Ackley}, {Adams}, {Adams}, {Addesso}, {Adhikari}, \& et~al.}]{gw1} {Abbott}, B.~P., {Abbott}, R., {Abbott}, T.~D., {et~al.} 2016{\natexlab{a}}, Physical Review Letters, 116, 241103

\bibitem[{{Abbott} {et~al.}(2016{\natexlab{b}}){Abbott}, {Abbott}, {Abbott}, {Abernathy}, {Acernese}, {Ackley}, {Adams}, {Adams}, {Addesso}, {Adhikari}, \& et~al.}]{gw2} {Abbott}, B.~P., {Abbott}, R., {Abbott}, T.~D., {et~al.} 2016{\natexlab{b}}, Physical Review Letters, 116, 061102

\bibitem[Abbot et al.(2017)]{gw3} Abbot B. P., et al., 2017, Phys. Rev. Lett., 118, 22

\bibitem[Abe(2010)]{worm} Abe F., 2010, ApJ, 725, 1

\bibitem[Asada(2017)]{asada} Asada, H., 2017, preprint(arXiv:1711.01730)

\bibitem[{{Amendola} {et~al.}(1999){Amendola}, {Frieman}, \& {Waga}}]{1999MNRAS.309..465A} {Amendola}, L., {Frieman}, J.~A., \& {Waga}, I. 1999, \mnras, 309, 465

\bibitem[Bannister et al.(2016)]{ESE3} Bannister K. W., et al., 2016, Science, 351, 6271, 354-356

\bibitem[Beloborodov(2017)]{magnetarFRB} Beloborodov A. M., 2017, preprint(arXiv: 1702.08644)

\bibitem[Bisnovatyi-Kogan \& Tsupko(2009)]{BKT09} Bisnovatyi-Kogan G. S., Tsupko O. Yu., 2009, Grav. \& Cos., 15, 1, 20-27

\bibitem[Bisnovatyi-Kogan \& Tsupko(2010)]{BKT10} Bisnovatyi-Kogan G. S., Tsupko O. Yu., 2010, MNRAS, 404, 4, 1790-1800

\bibitem[Bisnovatyi-Kogan \& Tsupko(2015)]{review} Bisnovatyi-Kogan G. S., Tsupko O. Yu., 2015, Plasma Phys. Reports, 41, 7, 562-581

\bibitem[{{Bolejko} {et~al.}(2013){Bolejko}, {Clarkson}, {Maartens}, {Bacon},
    {Meures}, \& {Beynon}}]{2013PhRvL.110b1302B} {Bolejko}, K., {Clarkson}, C., {Maartens}, R., {et~al.} 2013, Physical Review Letters, 110, 021302

\bibitem[Bozza \& Postiglione(2015)]{bp15} Bozza V., Postiglione A., 2015, JCAP, 6, 036

\bibitem[Chatterjee et al.(2017)]{extragalacticFRB} Chatterjee S., 2017, Nature, 541, 7635

\bibitem[Chen et al.(2015)]{voidLens1} Chen B., Kantowski R. \& Dai X., 2015, ApJ, 804, 130

\bibitem[Clegg et al.(1998)]{cleggFey} Clegg A. W., Fey A. L., Lazio T. J. W., 1998, ApJ, 496, 1, 253-266

\bibitem[Cognard et al.(1993)]{ESEIntro1} Cognard I., et al., 1993, Nature, 366, 6453, 320-322

\bibitem[Coles et al.(2015)]{coles2015} Coles W. A., et al., 2015, ApJ, 808, 2, 113

\bibitem[Coles et al.(2010)]{coles2010} Coles W. A., et al., 2010, ApJ, 717, 2, 1206-1221

\bibitem[Cordes \& Rickett(1998)]{cordesRickett1998} Cordes J. M., Rickett B. J., 1998, ApJ, 507, 2, 846-860

\bibitem[Cordes et al.(2017)]{FRBplasma1} Cordes J. M., Hessels J. W. T., Lazio T. J. W., Chatterjee S., Wharton R. S., 2017, ApJ, 842, 1

\bibitem[Cutler \& Thorne(2002)]{ELFGWaves} Cutler C., Thorne K. S., 2002, preprint (arXiv: gr-qc/0204090)

\bibitem[Deguchi \& Watson(1986)]{deguchiWatson1986b} Deguchi S., Watson W. D., 1986, Phys. Rev. D, 34, 1708-1718

\bibitem[Er \& Mao(2014)]{erMao} Er X., Mao S., 2014, MNRAS, 437, 3, 2180-2186

\bibitem[Fiedler et al.(1987)]{ESE0} Fiedler R. L., Dennison B., Johnston K. J., Hewish A., 1987, Nature, 326, 675-678

\bibitem[Fiedler et al.(1994)]{F94} Fiedler R., Dennison B., Johnston K.J., Waltman E.B., Simon R.S., 1994, ApJ, 430, 581

\bibitem[Goldreich \& Sridhar(2006)]{plasmaSheets} Goldreich P., Sridhar S., 2006, ApJ, 640, L159

\bibitem[Gordon(1928)]{gordon} Gordon W., 1928, Zeit. Phys., 48, 3-4, 180-191

\bibitem[Graham Smith et al.(2011)]{radioScatter} Graham Smith F., Lyne A. G., \& Jordan C., 2011, MNRAS, 410, 499

\bibitem[Ilic et al.(2013)]{ilic} Ilic S., Langer M., Douspis M., 2013, A\&A, 556, A51

\bibitem[Izumi et al.(2013)]{izumi} Izumi K., et al., 2013, Phys. Rev. D., 88, 2

\bibitem[Kantowski et al.(2013)]{embeddedLens} Kantowski R., Chen B., \& Dai X., 2013, Phys. Rev. D, 88, 083001

\bibitem[Kayser \& Schramm(1988)]{kayserSchramm} Kayser R., Schramm T., 1988, A\&A, 191, 1, 39

\bibitem[Kitamura et al.(2013)]{kna} Kitamura T., Nakajima K., Asada H., 2013, Phys. Rev. D., 87, 2

\bibitem[Krause et al.(2013)]{CMBLens} Krause, E., Chang, T.-C., Dor{\'e}, O., \& Umetsu, K. 2013, ApJ, 762, L20

\bibitem[Kulkarni \& Heiles(1988)]{ISMpressure} Kulkarni S. R., Heiles C., 1988, in Galactic and Extragalactic Radio Astronomy, eds. G. L. Verschuur \& K. I. Kellermann (Berlin : Springer), 95

\bibitem[Landau \& Lifshitz(1960)]{plasmaBook} Landau L.D., Lifshitz E.M., 1960, Electrodynamics of Continuous Media, Course of Theoretical Physics; Pergamon Press, Oxford, UK

\bibitem[Longo et al.(2006)]{starClusterLens} Longo P. et al., 2006, preprint (arXiv:astro-ph/0611551)

\bibitem[Narayan \& Bartelmann(1995)]{narayan} Narayan R., Bartelmann M., 1995, Proc. 1995 Jerusalem Winter School, eds. Dekel A., Ostriker J. P., Cambridge University Press

\bibitem[Pen \& King(2012)]{lens1} Pen U.-L., King L., 2012, MNRAS, 421, 1, L132-L136

\bibitem[Pen \& Levin(2014)]{ESE2} Pen U.-L., Levin Y., 2014, MNRAS, 442, 4, 3338-3346

\bibitem[Perlick(2000)]{perlickBook} Perlick V., 2000, Ray Optics, Fermats Principle, and Applications to General Relativity. Springer-Verlag, Berlin

\bibitem[Perlick(2004)]{worm2} Perlick V., 2017, Phys. Rev. D 69, 064017

\bibitem[Peters(1974)]{peters} Peters P. C., 1974, Phys. Rev. D., 9, 8, 2207-2218

\bibitem[Pushkarev et al.(2013)]{ESE1} Pushkarev A. B., et al., 2013, A\&A, 555, A80

\bibitem[Romani et al.(1987)]{romani87} Romani R. W., Blandford R. D., Cordes J. M., 1987, Nature, 328, 324-326

\bibitem[Rogers(2015)]{rogers15} Rogers A., 2015, MNRAS, 451, 1, 17-25

\bibitem[Rogers(2017a)]{rogers17a} Rogers A., 2017, MNRAS, 465, 2, 2151-2159

\bibitem[Rogers(2017b)]{rogers17b} Rogers A., 2017, Universe, 3, 1, 3

\bibitem[Safonova \& Torres(2002)]{saf2} Safonova M., Torres D. F., 2002, Mod. Phys. Lett. A., 17, 26

\bibitem[Safonova et al.(2002)]{safonova} Safonova M., Torres D. F., Romero G. E., 2002, Phys. Rev. D., 65, 2

\bibitem[Schneider, Ehlers \& Falco(1992)]{SEF} Schneider, P., Ehlers, J., \& Falco, E. E. 1992, Gravitational Lenses, ed. P. Schneider, J. Ehlers, \& E. E. Falco

\bibitem[Schramm \& Kayser(1987)]{schrammKayser} Schramm T., Kayser R., 1987, A\&A, 174, 1, 361

\bibitem[Selmke et al.(2011)]{selmke2} Selmke M., Braun M. \& Cichos F., 2011, ACS Nano, 6(3), 2741

\bibitem[Selmke et al.(2012)]{selmke3} Selmke M., Braun M., \& Cichos F., 2012, Opt. Express, 20(7), 8055

\bibitem[Selmke \& Cichos(2013)]{selmke} Selmke M., Cichos F., 2013, AJP, 81, 6, 405-413

\bibitem[Shannon \& Cordes(2017)]{shannonCordes2017} Shannon R. M., Cordes J. M., 2017, MNRAS, 464, 2, 2075-2089

\bibitem[Simard \& Pen(2017)]{ESE5} Simard D. \& Pen, U.-L., 2017, preprint(arXiv:1703.06855)

\bibitem[Spitler et al.(2016)]{repeatingFRB} Spitler L. G., et al., 2016, Nature, 531, 7593

\bibitem[Stinebring et al.(2001)]{stinebring} Stinebring D. R. et al., 2001, ApJ, 549, L97

\bibitem[Takahashi(2004)]{takahashi2} Takahashi R., 2004, A\&A, 423, 787-792

\bibitem[Takahashi(2017)]{takahashi3} Takahashi R., 2017, ApJ, 835, 1, 103

\bibitem[Tuntsov et al.(2016)]{ESE4} Tuntsov A. V., et al., 2016, ApJ, 817, 2,
176

\bibitem[Tuntsov et al.(2017)]{tuntsov17} Tuntsov A. V., et al., 2017, MNRAS, 469, 4, 5023

\bibitem[Vedantham et al.(2017a)]{newESE2017} Vedantham H. K., et al., 2017, preprint(arXiv:1702.05519)

\bibitem[Vedantham et al.(2017b)]{SAV} Vedantham H. K., et al., 2017, preprint(arXiv:1702.06582)

\bibitem[Virbhadra et al.(1998)]{virb} Virbhadra K.S., Narashima D., Chitre S.M., 1998, A\&A, 337, 1

\end{thebibliography}
\end{document}